\documentstyle[emulateapj,amsfonts,psfig,natbib,amsmath,rotating]{article}
\def\simlt{\mathrel{\hbox{\rlap{\hbox{\lower4pt\hbox{$\sim$}}}\hbox{$<$}}}}
\def\simgt{\mathrel{\hbox{\rlap{\hbox{\lower4pt\hbox{$\sim$}}}\hbox{$>$}}}}

\def\ale{\mathrel{\hbox{\rlap{\hbox{\lower4pt\hbox{$\sim$}}}\hbox{$<$}}}}
\def\age{\mathrel{\hbox{\rlap{\hbox{\lower4pt\hbox{$\sim$}}}\hbox{$>$}}}}

\def\nodata{---}

\def\cit{1}
\def\uh{2}
\def\ociw{3}
\def\pu{4}
\def\hf{5}
\def\srl{6}
\def\mil{7}
\def\vla{8}
\def\uva{9}
\def\anu{10}
\def\iac{11}
\def\llnl{12}
\def\ctio{13}
\def\gemini{14}

\begin{document}

\title{An {\it HST} Study of the Supernovae Accompanying GRB\,040924 and GRB\,041006}

\author{
A.~M. Soderberg\altaffilmark{\cit},
S.~R. Kulkarni\altaffilmark{\cit}, 
P.~A. Price\altaffilmark{\uh},
D.~B. Fox\altaffilmark{\cit},
E. Berger\altaffilmark{\ociw,\pu,\hf},
D.-S. Moon\altaffilmark{\srl,\mil},
S.~B. Cenko\altaffilmark{\srl},
A. Gal-Yam\altaffilmark{\cit,\hf},
D.~A. Frail\altaffilmark{\vla},
R.~A. Chevalier\altaffilmark{\uva},
L. Cowie\altaffilmark{\uh},
G.~S. Da Costa\altaffilmark{\anu},
A. MacFadyen\altaffilmark{\cit},
P.~J. McCarthy\altaffilmark{\ociw},
N. Noel\altaffilmark{\iac},
H.~S. Park\altaffilmark{\llnl},
B.~A. Peterson\altaffilmark{\anu},
M.~M. Phillips\altaffilmark{\ociw},
M. Rauch\altaffilmark{\ociw},
A. Rest\altaffilmark{\ctio},
J. Rich\altaffilmark{\anu},
K. Roth\altaffilmark{\gemini},
M. Roth\altaffilmark{\ociw},
B.~P. Schmidt\altaffilmark{\anu},
R.~C. Smith\altaffilmark{\ctio},
P.~R. Wood\altaffilmark{\anu},
}

\altaffiltext{\cit}{Division of Physics, Mathematics and Astronomy,
        105-24, California Institute of Technology, Pasadena, CA
        91125}
\altaffiltext{\uh}{University of Hawaii, Institute of Astronomy, 2680 Woodlawn Drive, Honolulu, HI 96822-1897}
\altaffiltext{\ociw}{Observatories of the Carnegie Institution of Washington,
        813 Santa Barbara St., Pasadena, CA 91101}
\altaffiltext{\pu}{Department of Astrophysical Sciences, Princeton University, 
        Princeton, NJ 08544}
\altaffiltext{\hf}{Hubble Fellow}
\altaffiltext{\srl}{Space Radiation Laboratory 220-47, California Institute of Technology, Pasadena, CA 91125}
\altaffiltext{\mil}{Robert A. Millikan Fellow}
\altaffiltext{\vla}{National Radio Astronomy Observatory, Socorro, NM 87801}
\altaffiltext{\uva}{Department of Astronomy, University of Virginia, P.O. Box 3818, Charlottesville, VA 22903-0818}
\altaffiltext{\anu}{Research School of Astronomy and Astrophysics, The Australian National University, Weston Creek, ACT 2611, Australia}
\altaffiltext{\iac}{Instituto de Astrofisica de Canarias, C/Via Lactea s/n, E-38200, La Laguna, Tenerife, Spain}
\altaffiltext{\llnl}{Lawrence Livermore National Laboratory, 7000 East Avenue, Livermore, CA 94550}
\altaffiltext{\ctio}{Cerro Tololo Inter-American Observatory, Casilla 603, La Serena, Chile}
\altaffiltext{\gemini}{Gemini Observatory, 670 N. Aohoku Place, Hilo, HI 96720}

\begin{abstract}

We present the results from a {\it Hubble Space Telescope/ACS} study
of the supernovae associated with gamma-ray bursts 040924 ($z=0.86$)
and 041006 ($z=0.71$).  We find evidence that both GRBs were
associated with a SN\,1998bw-like supernova dimmed by $\sim\,1.5$ and
$\sim\,0.3$ magnitudes, respectively, making GRB\,040924 the faintest
GRB-associated SN ever detected.  We study the luminosity dispersion
in GRB/XRF-associated SNe and compare to local Type Ibc supernovae
from the literature.  We find significant overlap between the two
samples, suggesting that GRB/XRF-associated SNe are not necessarily
more luminous nor produce more $^{56}$Ni than local SNe.  Based on the
current (limited) datasets, we find that the two samples may share a
similar $^{56}$Ni production mechanism.

\end{abstract}


\section{Introduction}

Gamma-ray burst (GRB) explosions harbor both spherical supernova (SN)
ejecta and highly collimated engine-driven jets
\citep{gvv+98,mgs+03,hsm+03}.  This ``spherical+jet'' paradigm for the
geometry of GRBs implies that both explosion components must be
studied independently.  Early ($t\lesssim$ few days) optical
observations trace the synchrotron radiation produced from the
engine-driven relativistic jets, while late-time ($t\gtrsim 20$ days)
observations probe the optical emission from the non-relativistic
spherical SN ejecta, powered by the decay of $^{56}$Ni.  Such
observations allow us to address the following fundamental questions
regarding the GRB/SN connection: what is the diversity among
GRB-associated SNe, and how does the sample compare to local Type Ibc
supernovae (SNe Ibc)?

Three of the best studied events, SNe 1998bw, 2003dh 2003lw (associated with
GRBs 980425, 030329 and 031203, respectively) were strikingly similar, with
brighter optical luminosities and faster photospheric velocities than local
SNe Ibc \citep{gvv+98,pcd+01,mgs+03,hsm+03,mtc+04}.  These events suggested
that the supernovae associated with GRBs belong to a distinct
sub-class of ``hyper-energetic'' supernovae.  However, there
are also several counter-examples: under-luminous SNe associated with GRBs
and X-ray flashes (XRFs; events very similar to GRBs but
differentiated by a soft prompt energy release peaking in the X-ray
band; \citealt{hik+01}) with optical luminosities significantly
fainter than SN\,1998bw.  The most extreme examples are GRB\,010921
(\citealt{pks+03}; but see \citealt{zkh04}) and XRF\,040701
\citep{skf+05}, for which the associated SNe are at least $1.3$
and $3.2$ magnitudes fainter than SN\,1998bw, respectively.

Motivated by our fundamental questions, we began a {\it Hubble Space
  Telescope} ({\it HST}) program to study the diversity in supernovae
  associated with GRBs and XRFs (GO-10135; PI: Kulkarni).  In
  \citet{skf+05}, we presented the results of our {\it HST} supernova
  search in XRFs 020903, 040701, 040812 and 040916.  We showed that at
  least some XRFs are associated with SNe, but that there may be a
  significant dispersion in their optical peak magnitudes.

Here we present the results from our {\it HST} study of the supernovae
associated with GRBs 040924 ($z=0.86$) and 041006 ($z=0.71$).  In
section \S\ref{sec:targets} we show that each is associated with a
SN\,1998bw-like supernova dimmed by $\sim\,1.5$ and $\sim\,0.3$
magnitudes, respectively, making GRB\,040924 the faintest
GRB-associated SN detected to date.  These two events suggest a
notable dispersion of peak optical magnitudes for GRB-associated SNe.

Finally, we compile optical peak magnitudes and $^{56}$Ni mass
estimates for all GRB/XRF-associated SNe and local SNe Ibc to study
the diversity among these two samples.  In \S\ref{sec:summary} we
address the question of whether both samples are drawn from the same
parent population of core-collapse supernovae.

\section{Supernova Light-Curve Synthesis}
\label{sec:templates}

In modeling the GRB-associated SNe, we adopted optical data for the
local SNe 1994I \citep{rvh+96}, 1998bw \citep{gvv+98,ms99} and 2002ap
\citep{fps+03a} as templates.  These three SNe were selected based on
their well-sampled optical light-curves which represent an overall
spread in the observed properties of Type Ibc supernovae.  To produce
synthesized light-curves for each of these template SNe, we compiled
optical $UBVRI$ observations from the literature and smoothed the
extinction-corrected (foreground plus host galaxy) light-curves.  We
then redshifted the light-curves by interpolating over the photometric
spectrum and stretching the arrival time of the photons by a factor of
$(1+z)$.  The details of each template dataset are described in
\citet{skf+05}.

\section{Hubble Space Telescope GRB-SN Search}
\label{sec:targets}

Using the {\it Wide-Field Camera} ({\it WFC}) of the {\it Advanced
  Camera for Surveys} ({\it ACS}) on-board {\it HST}, we imaged the
  fields of GRBs 040924 and 041006.  For each target we undertook
  observations at several epochs, spanning $t\sim 30$ and $\sim 150$
  days, in order to search for optical emission associated with an
  underlying supernova.  Each epoch consisted of 2 to 4 orbits during
  which we imaged the field in filters, F775W and/or F850LP,
  corresponding to SDSS $i'$- and $z'$- bands, respectively.

The {\it HST} data were processed using the {\tt multidrizzle} routine
\citep{fh02} within the {\tt stsdas} package of IRAF.  Images were
drizzled using {\tt pixfrac}=0.8 and {\tt pixscale}=1.0 resulting in a
final pixel scale of 0.05 arcsec/pixel.  Drizzled images were then
registered to the first epoch using the {\tt xregister} package within
IRAF.

To search for source variability and remove host galaxy contamination,
we used the {\it ISIS} subtraction routine by \citet{a00} which
accounts for temporal variations in the stellar PSF.  Adopting the
final epoch observations as template images, we produced residual
images. These residual images were examined for positive sources
positionally coincident with the afterglow error circle.  To test our
efficiency at recovering faint transient sources, false stars with a
range of magnitudes were inserted into the first epoch images using
{\it IRAF} task {\tt mkobject}.  The false stars were overlayed on top
of the diffuse host galaxy emission at radial distances similar to
that measured for the afterglow. An examination of the false stellar
residuals provided an estimate of the magnitude limit ($3\sigma$) to
which we could reliably recover faint transients.

Photometry was performed on the residual sources within a 0.5 arcsec
aperture.  We converted the photometric measurements to infinite
aperture and calculated the corresponding AB magnitudes within the native
{\it HST} filters using the aperture corrections and zero-points
provided by \citet{sjb+05}.  For comparison with ground-based data, we
also converted the photometric measurements to Johnson 
$I$- and $z$-band (Vega) magnitudes using the transformation coefficients
derived by \citet{sjb+05} and adopting the source color given by the 
first epoch F775W and F850LP observations.

In the following sections we summarize the afterglow properties for
both targets and the photometry derived from our {\it HST}
SN study.  A log of the {\it HST} observations for the GRBs follows in
Table~\ref{tab:hst}.

\subsection{GRB\,040924} 
\label{sec:grb040924}

\subsubsection{Prompt Emission and Afterglow Properties}
\label{sec:ag_040924}

GRB\,040924 was detected by the Wide-Field X-ray Monitor ({\it WXM})
on-board the High Energy Transient Explorer ({\it HETE-2}) satellite
on 2004 September 24.4951 UT.  Preliminary analysis showed the peak of
the spectral energy distribution was soft, $E_{\rm peak}\approx 42\pm
6$ keV, and the ratio of X-ray ($7-30$ keV) to $\gamma-$ray ($30-400$
keV) fluence, $S_X/S_{\gamma}\approx 0.6$, classifying this event as
an X-ray rich burst \citep{fra+04}.  Using the Roboticized Palomar
60-inch telescope ({\it P60}; Cenko {\it et al.}, in prep), we
discovered the optical afterglow at position $\alpha=02^{\rm h}
06^{\rm m} 22^{\rm s}.55$, $\delta=+16^{\circ} 06' 48''.8$ (J2000)
with an uncertainty of 0.2 arcsec in each coordinate \citep{f04}, well
within the 6.4 arcsec (radius) localization region.  As shown by our
extensive {\it P60} monitoring, the afterglow was $R\approx 18.0$ mag
at $t\approx 16$ minutes and subsequently decayed.

\begin{figure}
\vspace{-1.5cm}
\plotone{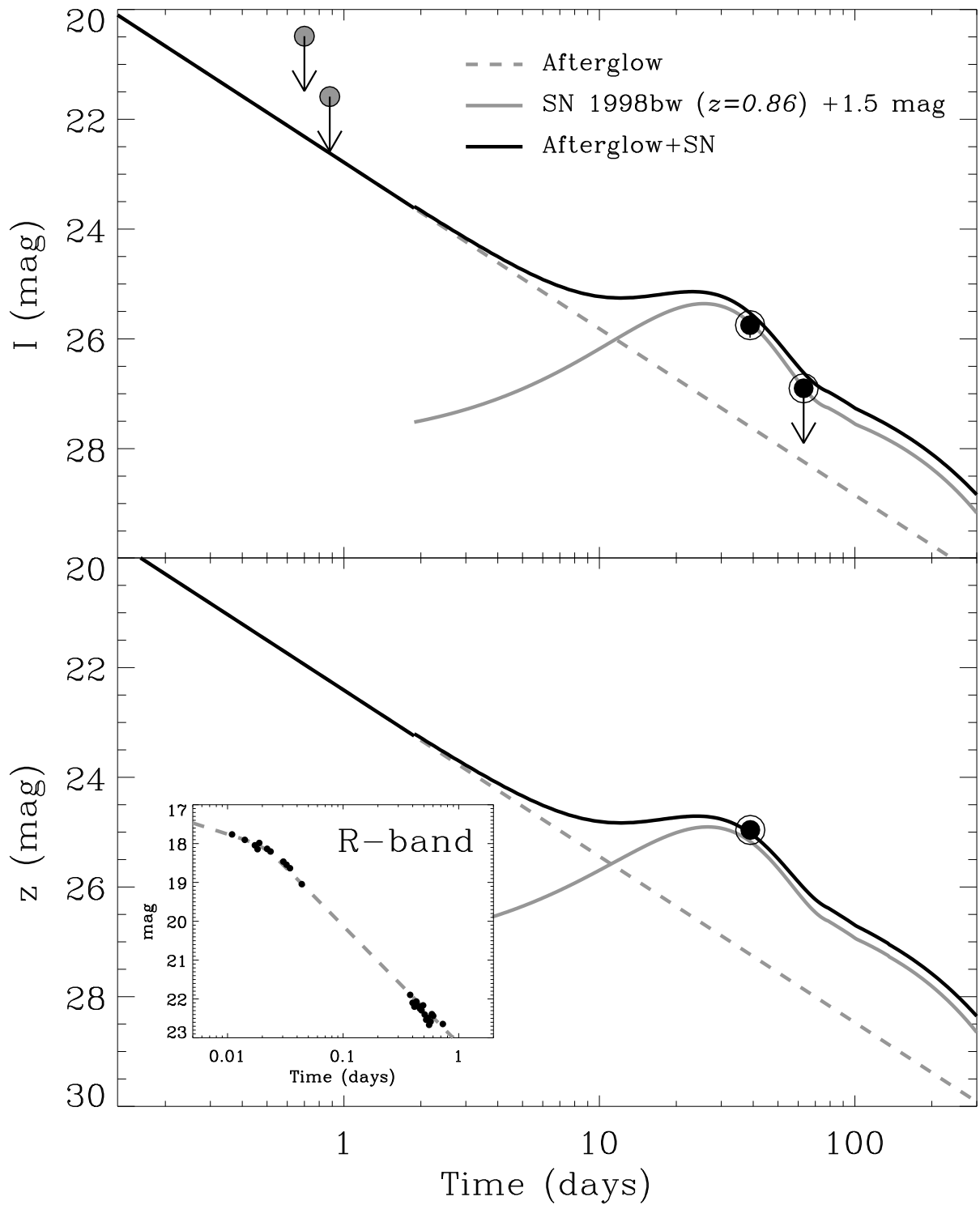}
\vspace{-2cm}
\caption{Constraints on a supernova associated with GRB\,040924.  As
  discussed in \S\ref{sec:ag_040924}, we fit a broken-power law model
  to our Palomar 60-inch $R$-band afterglow data reported in
  Table~\ref{tab:GRB040924} (inset). Adopting a spectral index of
  $\beta\approx -0.7$ we then extrapolate the $R$-band afterglow fit
  to the $I$- and $z$-bands (grey dashed lines).  Extinction-corrected
  data have been compiled from the GCNs and are overplotted (grey
  arrows; \citealt{sab+04}).  We fit the late-time {\it HST} data
  (encircled black dots) by summing the contribution from the
  afterglow model plus that from a supernova.  We find a best fit
  (solid black lines) by including a SN\,1998bw-like supernova (grey
  solid lines) redshifted to $z=0.86$ and dimmed by $\sim 1.5$
  magnitudes.
\label{fig:GRB040924_SN_curve}}
\end{figure}

We continued to monitor the afterglow emission from $t\approx 0.01$ to
1 day after the burst (Table~\ref{tab:GRB040924}), producing a
well-sampled $R$-band light-curve.  As shown in
Figure~\ref{fig:GRB040924_SN_curve}, these data show a shallow initial
decay, followed by a steepening at $t\sim 0.02$ days. In an effort to
characterize the light-curve behavior, we fit a smoothed, broken
power-law model (e.g. \citealt{bhr+99}) of the form

\begin{equation}
F_{\nu}(t) = 2 F_{\nu,0}\left[\left(\frac{t}{t_b}\right)^{\alpha_1 s} + \left(\frac{t}{t_b}\right)^{\alpha_2 s}\right]^{-1/s}
\end{equation}

\noindent
where $t_b$ is the break time in days, $F_{\nu,0}$ is the normalized
flux density at $t=t_b$, and $\alpha_1$ and $\alpha_2$ are the
asymptotic indices at $t \ll t_b$ and $t \gg t_b$, respectively.  Here
$s$ is used to parameterize the sharpness of the break.  A best
fit ($\chi^2_r=3.3$) is found for the following parameters:
$F_{\nu,0}\approx 90.4~\mu$Jy, $\alpha_1\approx 0.39$,
$\alpha_2\approx 1.22$, $t_b\approx 0.021$ days, and $s\approx 10$.
To interpret this steepening as a jet break \citep{r99,sph99} would
imply a electron spectral index, $N\propto \gamma^{-p}$ with $p\approx
1.2$.  This value is significantly lower than those typically observed
($p\approx 2.2$; \citealt{yhs+03}) and far below the range predicted
by the standard blastwave model, $p=[2-3]$ \citep{spn98}.  We
therefore ascribe this early afterglow phase (and steepening) to
another process, perhaps similar to that observed for the initial slow
decay of GRB\,021004 \citep{fyk+03}, interpreted as interaction 
with a stellar wind medium \citep{lc03}.

\citet{ta04} report that the afterglow was $K\approx 17.5\pm 0.1$ mag at
$t\approx 0.1$ days.  Comparison with the $R$-band afterglow model at
the same epoch provides a spectral index, $F_{\nu}\propto \nu^{\beta}$
with $\beta\approx -0.7$, between the $R$- and $K$-bands.  Adopting
this spectral index, we interpolate our $R$-band afterglow model to
the $I$- and $z$-bands for comparison with our late-time {\it HST}
observations (\S~\ref{sec:hst_040924}).

Optical spectroscopy later revealed that the burst was located at a
redshift of $z=0.859$, based on identification of several host galaxy
emission lines \citep{wsr+04}.

\subsubsection{{\it HST} observations}
\label{sec:hst_040924}

{\it HST/ACS} imaging was carried out on 2004 November 2.4, 26.5 and
2005 February 19.0 UT ($t \approx 39$, 63 and 147 days after the
burst).  For the first and third epochs, we observed with both the
F775W and F850LP filters, while the second epoch consisted of just
F775W observations.  We astrometrically tied the {\it HST} and {\it
  P60} images using four sources in common resulting in a final
systematic uncertainty of 0.22 arcsec ($2\sigma$).

\begin{figure}
\vspace{-2cm}
\plotone{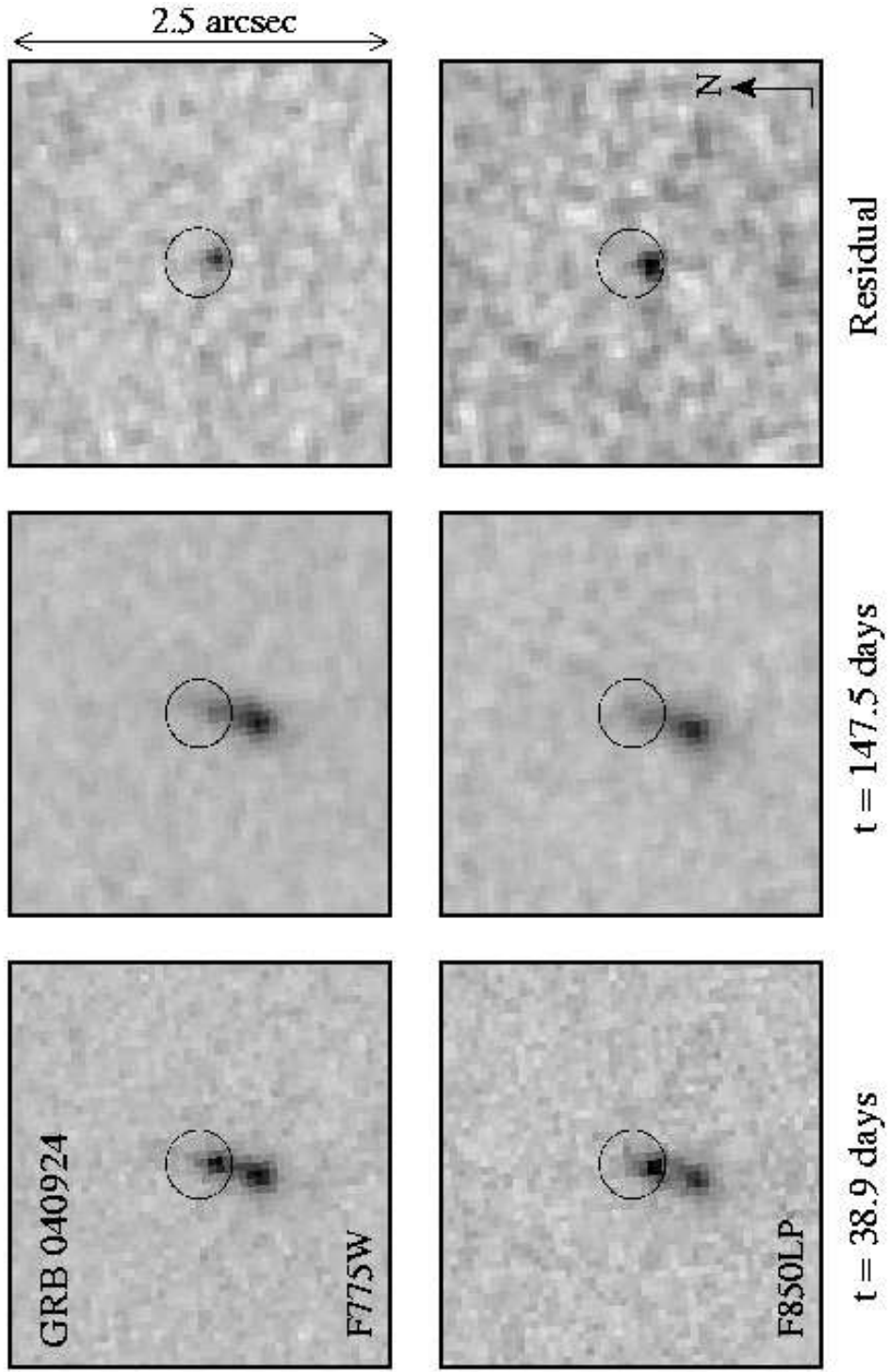}
\vspace{1cm}
\caption{Six-panel frame showing {\it HST/ACS} imaging for GRB\,040924
  at $t\sim 38.9$ days (Epoch 1) and $t\sim 147.5$ days (Epoch 3) in
  the F775W and F850LP filters. By subtracting Epoch 3 images from
  those of Epoch 1, we produced the residual images shown above.  We
  have applied the same stretch to all frames. As clearly shown in the
  residual images, a transient source is detected coincident with the
  $0.22$ arcsec (2$\sigma$) optical afterglow position (circle).
\label{fig:GRB040924_HST}}
\end{figure}

Our {\it HST} observations reveal that the optical afterglow error circle
coincides with a strongly variable source at $\alpha=02^{\rm h}
06^{\rm m} 22^{\rm s}.552$, $\delta=+16^{\circ} 06' 49''.11$ (J2000),
lying $\sim 0.3$ arcsec NW of the host galaxy nucleus.  We interpret this
source as a combination of afterglow plus supernova emission.
Figure~\ref{fig:GRB040924_HST} shows the first epoch, template image
and first epoch residual for both the F775W and F850LP filter
observations.  There is no evidence for a point source in
our third epoch (template) images.  We therefore make the reasonable
assumption that the transient source flux is negligible at that time.
The resulting {\it HST} photometry is listed in Table~\ref{tab:hst}.

Figure~\ref{fig:GRB040924_SN_curve} shows the {\it HST} photometry
along with our $I$- and $z$-band afterglow models (see
\S\ref{sec:ag_040924}) and some early-time data compiled from the GCNs
\citep{sab+04}, all corrected for Galactic extinction.  By comparing
the afterglow models with the {\it HST} photometry, we find
evidence for a $\sim 1.9$ magnitude rebrightening at $t\sim 39$ days.
We note that the timescale of this rebrightening is roughly consistent
with the peak time of SN\,1998bw at $z=0.86$.

\subsubsection{Associated Supernova}
\label{sec:sn_040924}

We interpret the observed late-time rebrightening as an associated
supernova component.  Comparison with our synthesized (redshifted)
supernova light-curves reveals that an associated SN\,1998bw-like
supernova would be $\sim 1.5$ magnitudes brighter than the {\it HST}
observation at $t\sim 39$ days, while SN\,1994I and SN\,2002ap-like
light-curves are fainter by $\sim 0.8$ and $\sim 1.5$ magnitudes,
respectively.  To fit the {\it HST} observations, we add the
contribution from a template SN light-curve (see
\S\ref{sec:templates}) to the afterglow model.  We allow the
brightness of the synthesized template to be scaled but do not allow
for the light-curve to be stretched.  Moreover, we assume the GRB and
SN exploded at the same time.  By fitting the {\it HST} $I$- and
$z$-band data simultaneously, we find a best fit solution
($\chi^2_r\approx 0.8$) for a SN\,1998bw-like supernova dimmed by $\sim
1.5$ magnitudes (Figure~\ref{fig:GRB040924_SN_curve}).  We note that
on this timescale, the supernova emission clearly dominates that of
the afterglow model.  Therefore, even if the afterglow underwent a
late-time ($t > 2$ days) jet break, our supernova fits would not be
significantly affected.

Host galaxy extinction could affect our SN light-curve fits since it
implies the associated SN was more luminous than the
observations suggest.  Fortunately, we can constrain the host galaxy
extinction using optical/IR afterglow observations, since extinction
will produce an artificial steepening of the synchrotron spectrum.
Making the reasonable assumption that the synchrotron cooling frequency
was above the optical/IR bands at early-time, we have $\beta_{\rm
  opt/IR}\approx -0.7=-(p-1)/2$ and thus $p\approx 2.4$.  For $p<2$
the standard blastwave model is violated, therefore implying the
spectral index must be steeper than $\beta_{\rm opt/IR}=-0.5$.
Adopting this limit for the intrinsic spectral index of the afterglow
implies a constraint on the host-galaxy extinction of $A_{V,\rm
  host}\lesssim 0.16$ mag.  This corresponds to limits on the
observed extinction of $A_I\approx A_z\lesssim 0.20$ mag, implying 
the associated SN was at least $1.3$ magnitudes fainter than
SN\,1998bw.  We therefore conclude that the supernova associated with
GRB\,040924 was between 1.3 and 1.5 magnitudes fainter than SN\,1998bw
at maximum light, making this event the faintest GRB-associated SN
ever detected.

\subsection{GRB\,041006}
\label{sec:grb041006}

\subsubsection{Prompt Emission and Afterglow Properties}
\label{sec:ag_041006}

GRB\,041006 was discovered by the {\it HETE-2/WXM} on 2004 October 6.513
UT.  The ratio of $2-30$ keV and $30-400$ keV channel fluences showed
the event was an X-ray rich burst.  The event was localized to a 5.0 arcmin
(radius) localization region centered at $\alpha=00^{\rm h} 54^{\rm m}
53^{\rm s}$, $\delta=+01^{\circ} 12' 04''$ (J2000; \citealt{gra+04}).  

Using the Siding Springs Observatory ({\it SSO}) 40-inch telescope, we
discovered the optical afterglow at $\alpha=00^{\rm h} 54^{\rm m}
50^{\rm s}.17$, $\delta=+01^{\circ} 14' 07''.0$ (J2000;
\citealt{dnp04}), consistent with the {\it HETE-2} error circle.  As
our early time ($t\approx 0.024$ to 0.122 days;
Table~\ref{tab:grb041006_sso}) {\it SSO} 40-inch and 2.3 meter
observations show, the afterglow was $V\approx 18.1$ mag at $t\approx
35$ minutes and subsequently faded in the $B$-, $V$- and $H$-bands.

\citet{sgn+05} have presented an extensive compilation of
(extinction-corrected) $R$-band afterglow data from $t\approx 0.03$ to
64 days, primarily from the Multiple Mirror Telescope (MMT) and from
the GCNs (see references therein).  They fit the evolution of the
early light-curve (up to $t\sim 4$ days) with the broken power-law
model given in Equation 1, finding best fit parameters of
$F_{\nu,0}\approx 50.0~\mu$Jy, $\alpha_1\approx 0.57$,
$\alpha_2\approx 1.29$, $t_b\approx 0.14$ days, and $s\approx 2.59$.
As in the case of GRB\,040924 (\S~\ref{sec:ag_040924}), this observed
steepening is too shallow to be interpreted as a jet break under the
standard blastwave model.  We instead interpret this early afterglow
phase as due to another process, possibly the result of an off-axis
viewing angle as proposed by \citet{grp05}.

Combining our early-time {\it SSO} data with those compiled by
\citet{sgn+05}, we estimate the spectral index of the optical/IR
afterglow to be $\beta_{\rm opt/IR}\approx -0.5$ by fitting a
power-law to the $B$-, $V$-, $R$- and $H$-band data. Adopting this
spectral index, we interpolate the $R$-band model to the $I$- and
$z$-bands for comparison with our late-time {\it HST} observations
(\S~\ref{sec:hst_041006}).

\begin{figure}
\vspace{0cm}
\plotone{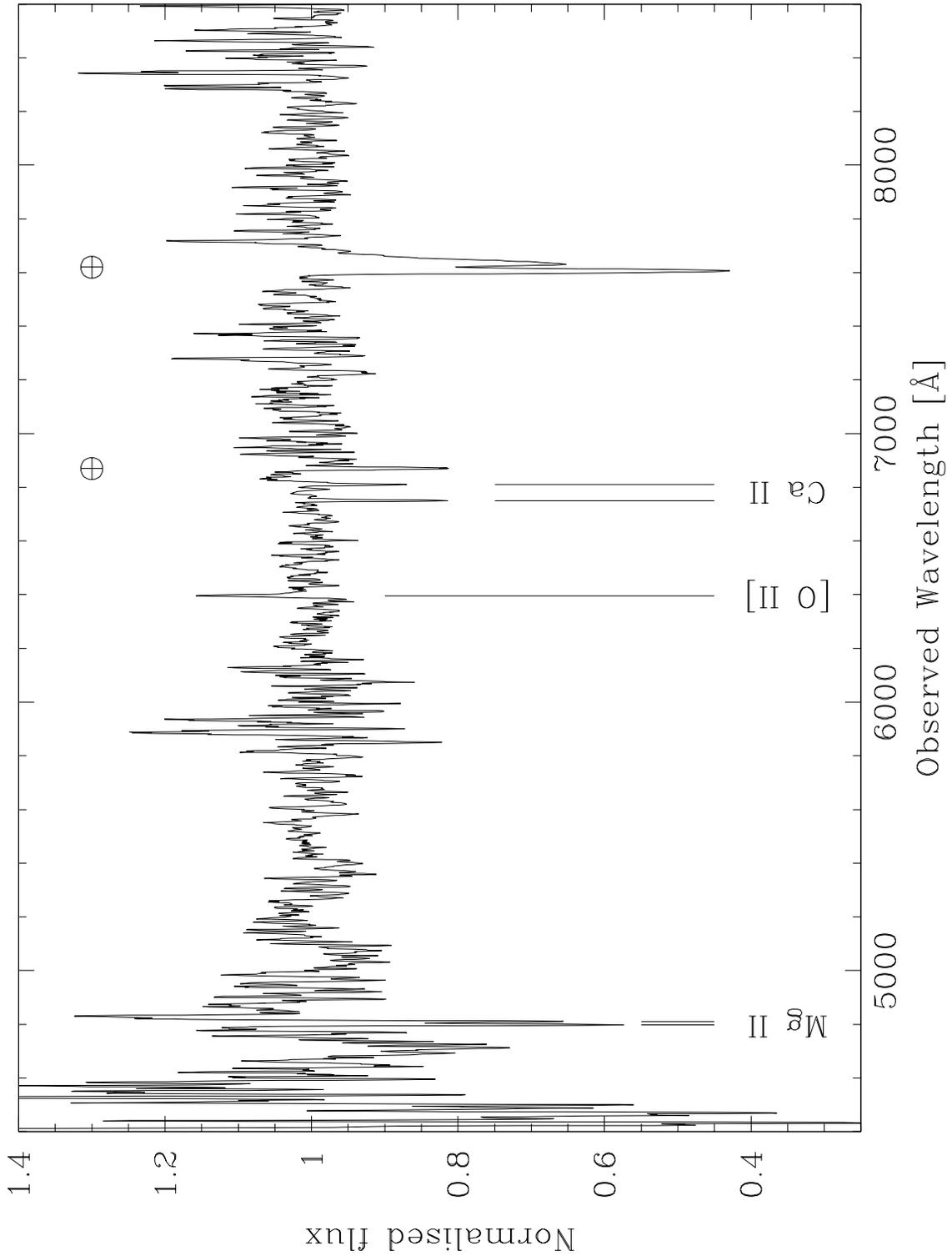}
\vspace{-2cm}
\caption{Gemini/GMOS spectrum of the optical afterglow for GRB\,041006
  at $t\approx 0.85$ days.  Based on three absorption features and one
  emission line, we determine the redshift of the host galaxy to be
  $z=0.716$, consistent with the earlier report of $z=0.712$ by
  \citet{ffc+04}.
\label{fig:grb041006_z}}
\end{figure}

Optical spectroscopy at $t\approx 0.6$ days revealed that the burst
was located at $z\ge 0.712$, based on tentative identification of several
absorption lines in the low signal-to-noise spectrum \citep{ffc+04}.
To confirm this redshift measurement, we acquired a spectrum of the
optical afterglow with the Gemini Multi-Object Spectrograph (GMOS) on
Gemini North (GN-2004B-Q-5; PI: Price) commencing at 2004 October 7.36
UT ($t\approx 0.85$ days).  Using the R400 grating
and a 1 arcsec slit, we obtained four individual 1800~sec exposures.
Data were reduced and extracted using the {\tt gemini.gmos} package
within IRAF.  We identify four absorption lines and one emission line
(Table \ref{tab:grb041006_spec} and Figure~\ref{fig:grb041006_z}),
which we interpret as arising from \ion{Mg}{2}, \ion{Ca}{2} and
[\ion{O}{2}] at a redshift of $z = 0.716$, in broad agreement with the
earlier measurement by \citet{ffc+04}.  Based on the high equivalent
widths and the presence of [\ion{O}{2}], this is likely the redshift
of the host galaxy, and not due to an intervening system along the
line of sight.  Using the $R$-band afterglow model and the
photometrically derived spectral index, we flux calibrate the spectrum
and obtain an approximate [\ion{O}{2}] emission line flux of $1.3
\times 10^{-17}$~erg/cm$^2$/s.  Adopting the conversion factor of
\citet{k98}, this line flux corresponds to a star formation rate of
0.5~M$_\odot$/yr at the redshift of the host galaxy.\footnote{We have
  adopted a cosmology with $H_0 = 65$~km/s/Mpc and
  $(\Omega_M,\Omega_\Lambda) = (0.3,0.7)$.}

\subsubsection{{\it HST} observations}
\label{sec:hst_041006}

{\it HST/ACS} imaging was carried out on 2004 November 2.3, 27.4,
December 23.4 and 2005 February 10.5 UT ($t \approx 27$, 52, 78 and
127 days after the burst).  For the first and fourth epochs, we
observed with both the F775W and F850LP filters, while the second and
third epochs consisted of just F775W observations.  We astrometrically
tied the {\it HST} and {\it SSO} images using five sources in common
resulting in a final systematic uncertainty of 0.53 arcsec
($2\sigma$).

Our {\it HST} observations reveal that the optical afterglow error circle
coincides with a strongly variable source at $\alpha=00^{\rm h}
54^{\rm m} 50^{\rm s}.229$, $\delta=+01^{\circ} 14' 05''.82$ (J2000).
The source is situated directly on top of a faint host galaxy which is
at the detection limit of our {\it HST} images.  We note the presence
of a brighter galaxy $\sim 1$ arcsec from the afterglow position which
was misidentified as the host galaxy by \citet{cmt+04} based on
ground-based images.

\begin{figure}
\vspace{-2cm}
\plotone{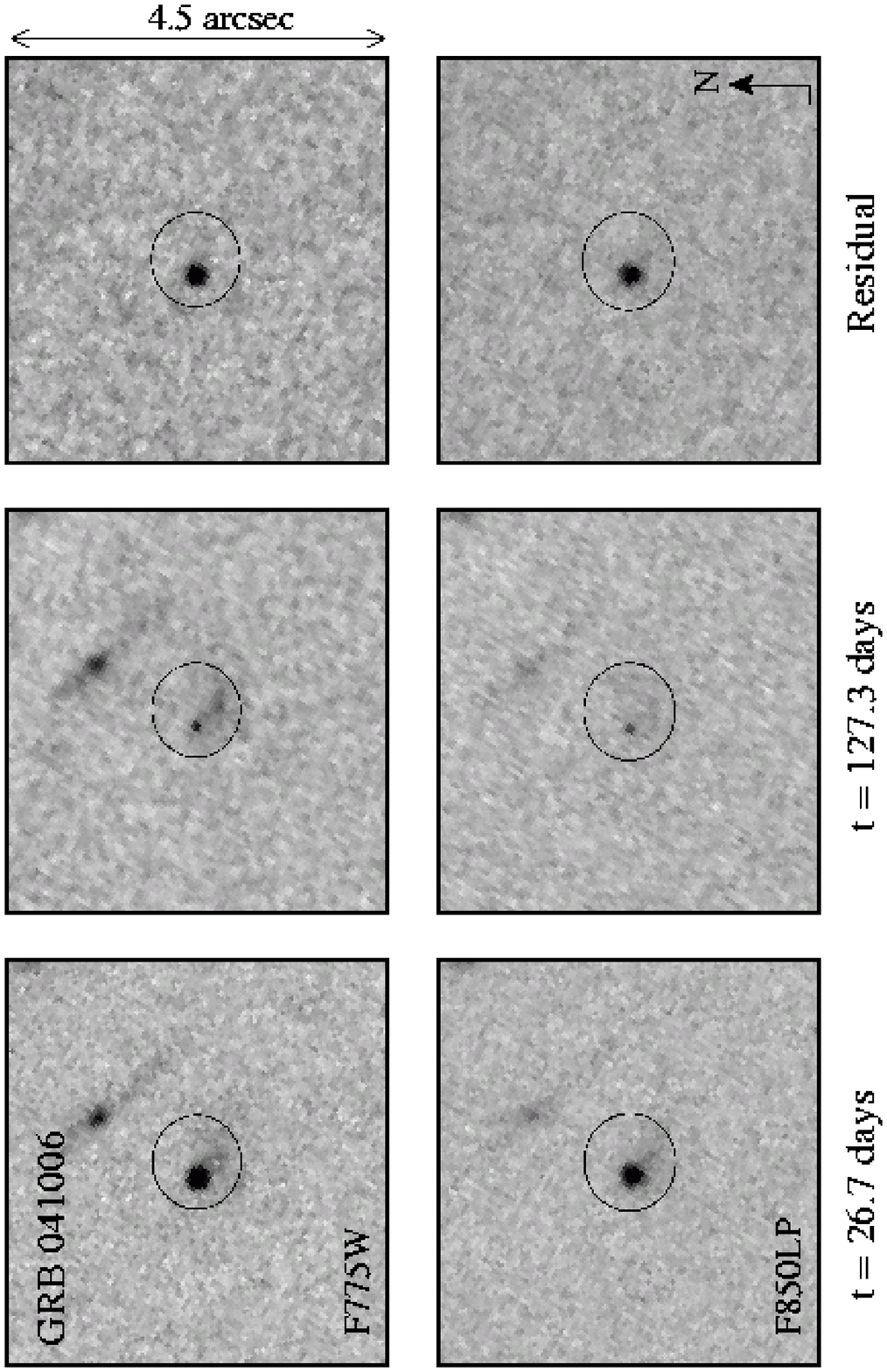}
\vspace{1cm}
\caption{Six-panel frame showing {\it HST/ACS} imaging for GRB\,041006
  at $t\sim 26.7$ days (Epoch 1) and $t\sim 127.3$ days (Epoch 4) in
  the F775W and F850LP filters. By subtracting Epoch 4 images from
  those of Epoch 1, we produced the residual images shown above.  We
  have applied the same stretch to all frames. As clearly shown in the
  residual images, a transient source is detected coincident with the
  $0.53$ arcsec (2$\sigma$) optical afterglow position (circle).
\label{fig:GRB041006_HST}}
\end{figure}

Figure~\ref{fig:GRB041006_HST} shows the first epoch, template image
and first epoch residual for both the F775W and F850LP filter
observations.  From the figure, it is clear that the source is still
faintly detected in our final epoch (template) observations.  To
assume negligible source flux in the template images is therefore not
valid. We estimate the flux of the source in our template images by
performing PSF photometry on the object.  In calculating the magnitude
of the source in the first, second and third epochs, we add the
template epoch flux to the residual image photometry.  The resulting
{\it HST} photometry is given in Table~\ref{tab:hst}.

\begin{figure}
\vspace{-1.5cm}
\plotone{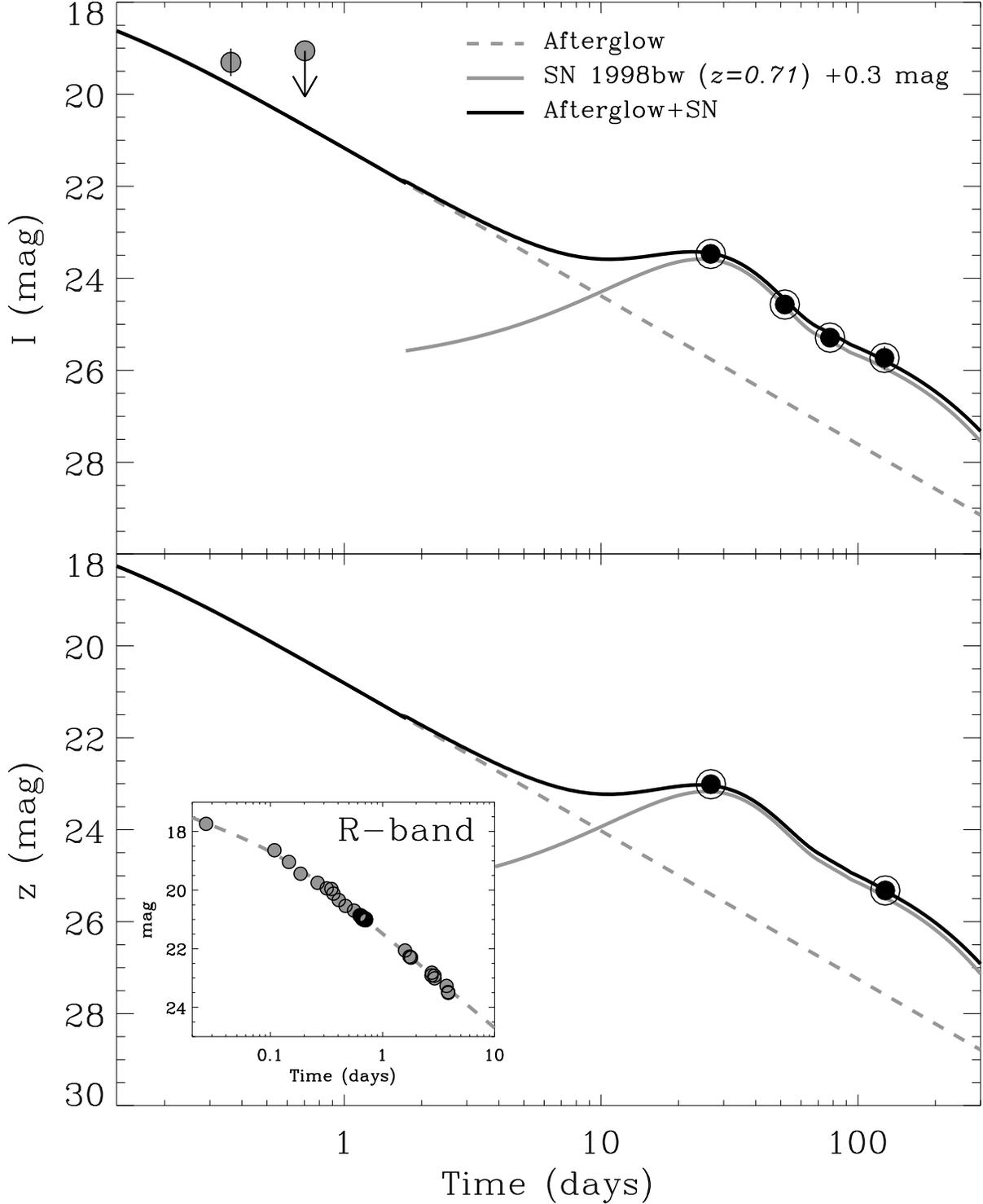}
\vspace{-2cm}
\caption{Constraints on a supernova associated with GRB\,041006.
  Adopting the $R$-band afterglow data reported in \citet{sgn+05}, we
  reproduce their broken-power law fit (inset).  Adopting a spectral
  index of $\beta\approx -0.5$ we then extrapolate the $R$-band
  afterglow fit to the $I$- and $z$-bands (grey dashed lines).
  Extinction-corrected data have been compiled from the GCNs are
  overplotted (grey circles; \citealt{fbc+04,hcn+04}).  We fit the
  late-time {\it HST} data (encircled black dots) by summing the
  contribution from the afterglow model plus that from a supernova.
  We find a best fit (solid black lines) by including a
  SN\,1998bw-like supernova (grey solid lines) redshifted to $z=0.71$
  and dimmed by $\sim 0.3$ magnitudes.
\label{fig:GRB041006_SN_curve}}
\end{figure}

Figure~\ref{fig:GRB041006_SN_curve} shows the {\it HST} photometry
along with our $I$- and $z$-band afterglow model (see
\S\ref{sec:ag_041006}) and some early-time data compiled from the GCNs
\citep{fbc+04,hcn+04}, all corrected for Galactic extinction.  By
comparing the afterglow model with the {\it HST} data, we find
evidence for a $\sim 2.1$ magnitude rebrightening at $t\sim 27$ days,
roughly consistent with the peak time of a SN\,1998bw-like supernova
at $z=0.71$.  We note that this rebrightening was discovered earlier
($t\sim 12$ days) using ground-based facilities and interpreted as the
emerging flux from an associated supernova
\citep{bsa+04,gsp+04,sgn+05}.

\subsubsection{Associated Supernova}
\label{sec:sn_041006}

As discussed in \S~\ref{sec:ag_041006}, \citet{sgn+05} presented an
extensive compilation of $R$-band observations which span the epoch of
rebrightening.  By removing the contribution from the afterglow, they
find evidence for an associated SN with a peak magnitude $\sim 0.1$
mag brighter than SN\,1998bw and with a light-curve stretched
by a factor of 1.35 in time.  

To confirm this result, we compared our redshifted template SN
light-curves with our late-time {\it HST} observations and found that
an associated SN\,1998bw-like supernova would be $\sim 0.2$ mag
brighter than the {\it HST} observation at $t\sim 27$ days, while
SN\,1994I- and SN\,2002ap-like supernova light-curves are fainter by
$\sim 1.4$ and $\sim 2.3$ magnitudes, respectively.  In an effort to
characterize the light-curve of the associated SN, we fit the {\it
HST} observations in a manner similar to that done for GRB\,040924
(\S~\ref{sec:sn_040924}).  By fitting the {\it HST} $I$- and $z$-band
data simultaneously, we find a best fit solution ($\chi^2_r\approx
0.4$) for a SN\,1998bw-like supernova dimmed by $\sim 0.3$ magnitudes.
We emphasize that this fit does not include any stretching of the
light-curve.  Moreover, we find no improvement in the quality of the
fit when template stretching is included.  This result is
inconsistent with the findings of \citet{sgn+05}.  We attribute this
discrepancy to contamination from the host and nearby galaxy (see
Figure~\ref{fig:GRB041006_HST}) which plagues the \citet{sgn+05}
late-time photometry.

We note that the presence of host galaxy extinction would imply that
the associated SN was brighter than our estimates.  We use the
optical/IR afterglow observations to constrain the host galaxy
extinction in a manner similar to that performed for GRB\,040924
(\S\ref{sec:sn_040924}).  However, in this case, the observed spectral
index ($\beta_{\rm opt/IR}\approx -0.5$) is comparable to the
theoretical limit and is therefore consistent with negligible host
galaxy extinction.  We conclude that the supernova associated with
GRB\,041006 was $\sim 0.3$ magnitudes fainter than SN\,1998bw at
maximum light and decayed at a comparable rate.

\section{Discussion}
\label{sec:summary}

In the sections above, we showed that GRBs 040924 and 041006 were each
associated with a supernova similar to SN\,1998bw but dimmed by
$\sim\,1.5$ and $\sim\,0.3$ magnitudes, respectively.  These two
events clearly show that there is a significant spread in the
luminosity of GRB-associated SNe, suggesting a dispersion in the
production of $^{56}$Ni.  Such a dispersion has already been observed
for local SNe Ibc \citep{rbc+02}.

To better study this dispersion, we compile optical peak magnitudes
(rest-frame) and $^{56}$Ni mass estimates for GRB/XRF-associated SNe
and local SNe Ibc from the literature (Tables~\ref{tab:grb_MV_Ni} and
\ref{tab:sn_MV_Ni}).  In this compilation we consider all the
GRB/XRF-associated SNe with spectroscopic redshifts of $z\lesssim 1$;
SN searches at higher redshifts are plagued by UV (rest-frame)
line-blanketing.  Since spectroscopic confirmation of the SN component
is only available for the four nearest events in this sample (GRBs
980425, 020903, 031203 and 030329; see Table~\ref{tab:grb_MV_Ni}),
photometric data is used to constrain the brightness of an associated
SN.  We note, however, that uncertainty associated with host galaxy
extinction and decomposition of the afterglow emission from the SN can
impose a non-negligible uncertainty in the estimated SN peak magnitude.
For the local sample, we include only the events with published
photometry.  The latter may impose a slight bias in favor of brighter
and peculiar SNe\footnote{Three of the thirteen local SNe Ibc included
in this compilation were published as ``luminous'' events.  By
removing them from the local sample, the probability that both
GRB/XRF-associated SNe and local SNe are drawn from the same parent
population is reduced to $\sim 50\%$.}.

\begin{figure}
\vspace{-1.5cm}
\plotone{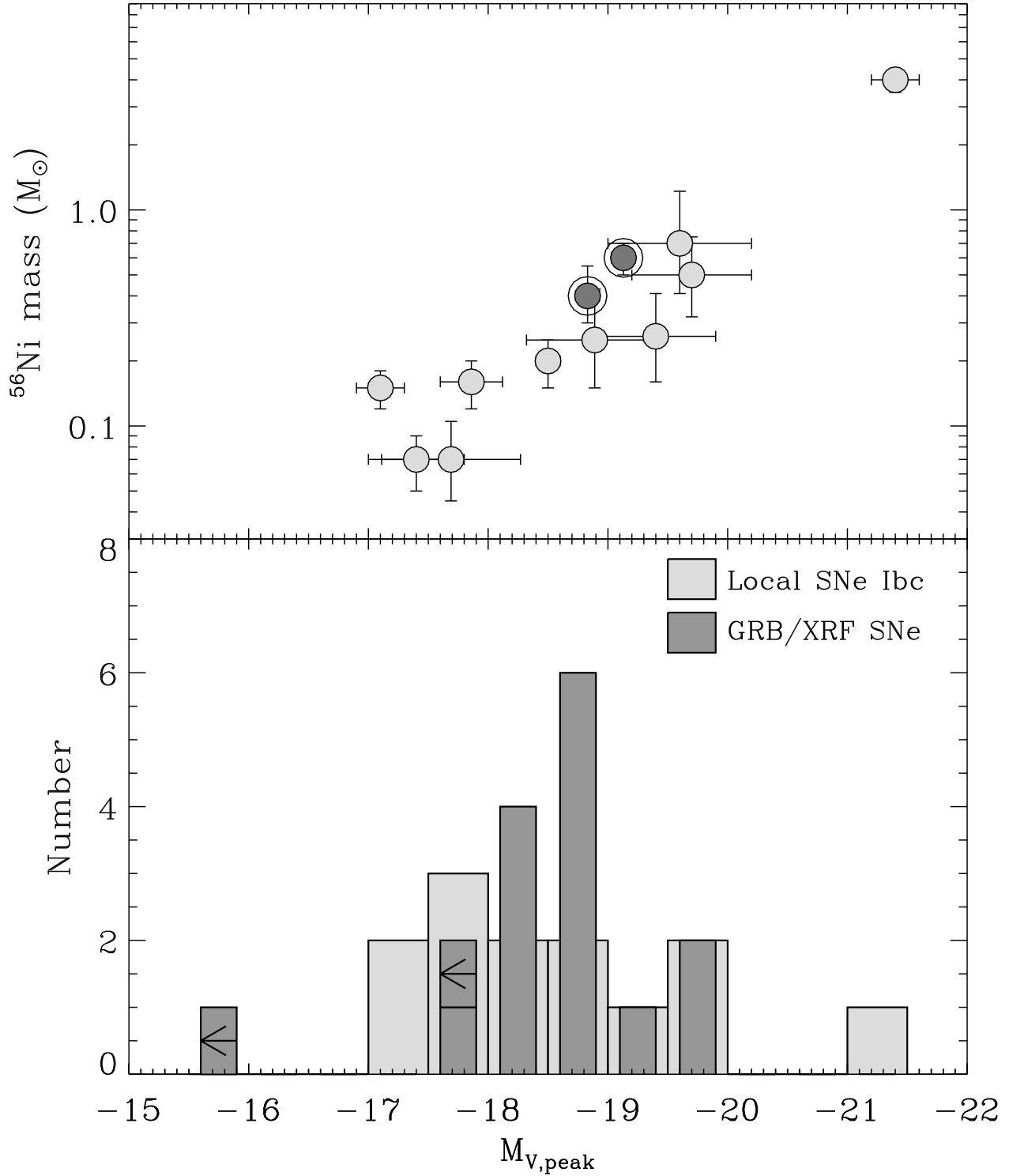}
\vspace{-2cm}
\caption{Top panel: rest-frame optical peak magnitudes for
  GRB-associated SNe (encircled dots) and local SNe Ibc (grey circles)
  have been compiled from the literature (Tables~\ref{tab:grb_MV_Ni}
  and \ref{tab:sn_MV_Ni}) and are plotted against the mass of
  $^{56}$Ni synthesized in the explosion.  For $^{56}$Ni estimates
  without errors, we adopt the fractional uncertainty associated with
  the peak luminosity. Peak optical magnitudes clearly trace the mass
  of $^{56}$Ni ejected.  Apparently, GRB-associated SNe do not
  necessarily produce more $^{56}$Ni than ordinary local SNe Ibc.
  Bottom panel: histogram of peak optical magnitudes for
  GRB/XRF-associated SNe and local SNe Ibc.  There is a significant
  overlap between the two samples.  Our K-S test reveals that there is
  a $\sim 91\%$ probability the two samples are drawn from the same
  population.
\label{fig:MV_Ni_hist}}
\end{figure}

In Figure~\ref{fig:MV_Ni_hist} (top panel), we show $^{56}$Ni mass
estimates and peak optical magnitudes for GRB/XRF-associated SNe and
local SNe Ibc.  From the figure, it is clear that (1) peak magnitude
scales roughly with $^{56}$Ni mass, and (2) GRB-associated SNe do not
necessarily synthesize more $^{56}$Ni than ordinary SNe Ibc.

We next compare their peak optical magnitudes, since the correlation
between $^{56}$Ni mass and peak optical magnitudes should manifest
itself as a similar dispersion.  In Figure~\ref{fig:MV_Ni_hist}
(bottom panel) we show a histogram of peak optical magnitudes
for all GRB/XRF-associated SNe and local SNe Ibc.  Examination of the
two samples reveals noticeable overlap; while GRB/XRF-associated SNe
tend to cluster toward the brighter end of the distribution, some of
the faintest SNe Ibc observed are in fact associated with GRBs and
XRFs (e.g. GRBs 010921, 040924, XRF\,040701) and the most luminous
events are actually drawn from the local sample.  Apparently,
GRB/XRF-associated SNe are not necessarily over-luminous in comparison
with local SNe Ibc.  Moreover, by assuming a rough scaling of
$^{56}$Ni production with peak optical magnitude, this implies a
notable overlap between the two samples with respect to their
dispersion in synthesized material, consistent with the data shown in 
the upper panel.

The following question naturally arises: are both GRB/XRF-associated
SNe and local SNe Ibc drawn from the same population of events?  To
address this issue, we performed a Kolmogorov-Smirnov (K-S) test on
the two sets of SN peak magnitudes.  Our K-S test reveals that the
probability the two samples have been drawn from the same parent
population is $\sim 91\%$.  It is therefore conceivable that both
SNe Ibc and GRB/XRF-associated SNe belong to the same SN
population, and thus share a common $^{56}$Ni production mechanism.

This result may imply that GRB/XRF-associated SNe and local SNe Ibc
share a common progenitor system and/or explosion mechanism.  Yet we
know from radio observations of local SNe that most Type Ibc's lack
the engine-driven relativistic ejecta observed in GRBs
\citep{bkf+03,sfw04,snk05}.  This may be accounted for in the spherical+jet
paradigm where the production of $^{56}$Ni and relativistic ejecta are
independent parameters of the explosion, each of which can be
individually tuned.  Additional studies of both GRB/XRF-associated SNe and
local SNe Ibc will enable us to better map out this
two-dimensional parameter space and address in detail the nature of the
progenitors and explosion mechanisms.

\begin{acknowledgements}
The authors are grateful for support under the Space Telescope Science
Institute grant HST-GO-10135.  A.M.S. acknowledges support by the NASA
Graduate Student Researchers Program. E.B. is supported by
NASA through Hubble Fellowship grant HST-HF-01171.01 awarded by the
STScI, which is operated by the Association of Universities for
Research in Astronomy, Inc., for NASA, under contract NAS 5-26555.
A.G. acknowledges support by NASA through Hubble Fellowship grant
\#HST-HF-01158.01-A awarded by STScI.  KR is supported by the Gemini
Observatory, which provided observations presented in this paper, and
which is operated by the Association of Universities for Research in
Astronomy, Inc., under a cooperative agreement with the NSF on behalf
of the Gemini partnership: the National Science Foundation (United
States), the Particle Physics and Astronomy Research Council (United
Kingdom), the National Research Council (Canada), CONICYT (Chile), the
Australian Research Council (Australia), CNPq (Brazil) and CONICET
(Argentina).

\end{acknowledgements}

\bibliographystyle{apj1b} 

\clearpage

\begin{deluxetable}{clrcrccc}
\tablecaption{{\it HST} Observation Log}
\tablewidth{0pt} \tablehead{ 
\colhead{Target} & \colhead{Date Obs} & \colhead{$\Delta t$} & 
\colhead{Exp. Time} & \colhead{Filter} & \colhead{{\it HST} mag\tablenotemark{a}} & 
\colhead{Extinction\tablenotemark{b}} & \colhead{Johnson mag\tablenotemark{c}} \\
\colhead{} & \colhead{(UT)} & \colhead{(days)} & 
\colhead{(sec)} & \colhead{} & \colhead{(AB)} & 
\colhead{$A_{\lambda}$} & \colhead{(Vega)} \\ } 

\startdata 
GRB\,040924 & 2004 November 2.4 & 38.9 & 3064 & F775W & $26.44\pm 0.23$ & $A_I=0.113$ & $I=25.75\pm 0.23$ \\ 
\nodata & 2004 November 2.5 & 39.0 & 3211 & F850LP & $25.60\pm 0.11$ & $A_z=0.106$ & $z=24.96\pm 0.11$ \\ 
\nodata & 2004 November 26.5 & 63.0 & 3932 & F775W & $>27.59$ & $A_I=0.113$ & $I>26.90$ \\ 
\nodata & 2005 February 18.5 & 147.0 & 3932 & F775W & \nodata & \nodata & \nodata \\ 
\nodata & 2005 February 19.5 & 148.0 & 3932 & F850LP & \nodata & \nodata & \nodata \\
GRB\,041006 & 2004 November 2.3 & 26.7 & 3832 & F775W & $24.01\pm 0.05$ & $A_I=0.044$ & $I=23.47\pm 0.05$ \\
\nodata & 2004 November 2.3 & 26.7 & 3430 & F850LP & $23.58\pm 0.04$ & $A_z=0.041$ & $z=23.01\pm 0.04$ \\
\nodata & 2004 November 27.4 & 51.9 & 4224 & F775W & $25.11\pm 0.10$ & $A_I=0.044$ & $I=24.57\pm 0.10$ \\
\nodata & 2004 December 23.4 & 77.8 & 4224 & F775W & $25.83\pm 0.19$ & $A_I=0.044$ & $I=25.29\pm 0.19$ \\
\nodata & 2005 February 10.3 & 126.8 & 4224 & F775W & $26.26\pm 0.25$ & $A_I=0.044$ & $I=25.73\pm 0.25$ \\
\nodata & 2005 February 11.2 & 127.7 & 4224 & F850LP & $25.90\pm 0.21$ & $A_z=0.041$ & $z=25.32\pm 0.21$ \\

\enddata 
\tablenotetext{a}{AB system magnitudes in the {\it HST}
  filters given in column 5.  Photometry was done on residual images (see \S\ref{sec:targets}).
  We have assumed the source flux to be negligible in the final
  (template) epoch for GRB\,040924.  For GRB\,041006, we estimated the
  source flux in the template epoch using PSF photometry and corrected
  the residual photometry accordingly.  All represent observed
  magnitudes, not corrected for foreground extinction.}
\tablenotetext{b}{Galactic extinction from \citet{sfd+98}.}
\tablenotetext{c}{Magnitudes from column 6, converted to the Vega
  system and corrected for foreground extinction using $A_{\lambda}$
  given in column 7.}
\label{tab:hst}
\end{deluxetable}

\clearpage

\begin{deluxetable}{rrr}
\tablecaption{Palomar 60-inch $R-$band Observations of
GRB\,040924} \tablewidth{0pt} \tablehead{ \colhead{Date Obs} &
\colhead{$\Delta t$} & \colhead{Magnitude\tablenotemark{a}} \\ 
\colhead{(UT)} & \colhead{(days)} & \colhead{} \\ }

\startdata 
2004 September 24.506 & 0.011 & $17.8\pm 0.1$ \\
2004 September 24.509 & 0.014 & $17.9\pm 0.1$ \\
2004 September 24.512 & 0.017 & $18.0\pm 0.1$ \\
2004 September 24.513 & 0.018 & $18.1\pm 0.1$ \\
2004 September 24.514 & 0.019 & $18.0\pm 0.1$ \\
2004 September 24.517 & 0.022 & $18.1\pm 0.1$ \\
2004 September 24.519 & 0.024 & $18.2\pm 0.1$ \\
2004 September 24.526 & 0.030 & $18.5\pm 0.1$ \\
2004 September 24.527 & 0.032 & $18.5\pm 0.1$ \\
2004 September 24.530 & 0.035 & $18.6\pm 0.1$ \\
2004 September 24.539 & 0.044 & $19.0\pm 0.1$ \\
2004 September 24.875 & 0.380 & $21.9\pm 0.1$ \\
2004 September 24.894 & 0.399 & $22.1\pm 0.1$ \\
2004 September 24.909 & 0.414 & $22.2\pm 0.1$ \\
2004 September 24.926 & 0.430 & $22.1\pm 0.1$ \\
2004 September 24.942 & 0.447 & $22.2\pm 0.1$ \\
2004 September 24.957 & 0.462 & $22.3\pm 0.1$ \\
2004 September 24.972 & 0.477 & $22.3\pm 0.1$ \\
2004 September 24.988 & 0.493 & $22.2\pm 0.1$ \\
2004 September 25.003 & 0.508 & $22.4\pm 0.1$ \\
2004 September 25.018 & 0.523 & $22.5\pm 0.1$ \\
2004 September 25.036 & 0.540 & $22.5\pm 0.1$ \\
2004 September 25.050 & 0.555 & $22.7\pm 0.1$ \\
2004 September 25.066 & 0.571 & $22.6\pm 0.1$ \\
2004 September 25.081 & 0.586 & $22.4\pm 0.1$ \\
2004 September 25.097 & 0.602 & $22.4\pm 0.1$ \\
2004 September 25.224 & 0.729 & $22.6\pm 0.1$ \\

\enddata \tablenotetext{a}{Corrected for Galactic extinction according to \citet{sfd+98}. }
\label{tab:GRB040924}
\end{deluxetable}

\clearpage

\begin{deluxetable}{rrcrr}
\tablecaption{{\it SSO} 40-inch and 2.3-meter Observations of GRB\,041006}
\tablewidth{0pt} \tablehead{ 
\colhead{Date Obs} & \colhead{$\Delta t$} & \colhead{Filter} & 
\colhead{Telescope} & \colhead{Magnitude\tablenotemark{a}} \\
\colhead{(UT)} & \colhead{(days)} & \colhead{} & \colhead{} & \colhead{} \\ }

\startdata 
2004 October 6.537 & 0.024 & $V$ & 40-inch & $18.11\pm 0.02$ \\
2004 October 6.541 & 0.028 & $V$ & 40-inch & $18.20\pm 0.04$ \\
2004 October 6.546 & 0.033 & $V$ & 40-inch & $18.29\pm 0.02$ \\
2004 October 6.550 & 0.037 & $V$ & 40-inch & $18.38\pm 0.03$ \\
2004 October 6.555 & 0.042 & $V$ & 40-inch & $18.48\pm 0.03$ \\
2004 October 6.582 & 0.069 & $B$ & 40-inch & $19.18\pm 0.07$ \\
2004 October 6.586 & 0.073 & $B$ & 40-inch & $19.24\pm 0.06$ \\
2004 October 6.590 & 0.077 & $B$ & 40-inch & $19.29\pm 0.07$ \\
2004 October 6.595 & 0.082 & $V$ & 40-inch & $18.99\pm 0.03$ \\
2004 October 6.605 & 0.092 & $H$ & 2.3-meter & $17.16\pm 0.07$ \\
2004 October 6.612 & 0.099 & $H$ & 2.3-meter & $17.12\pm 0.06$ \\
2004 October 6.620 & 0.107 & $H$ & 2.3-meter & $17.19\pm 0.06$ \\
2004 October 6.627 & 0.114 & $H$ & 2.3-meter & $17.47\pm 0.06$ \\
2004 October 6.635 & 0.122 & $H$ & 2.3-meter & $17.43\pm 0.07$ \\
\enddata \tablenotetext{a}{Corrected for Galactic extinction according to \citet{sfd+98}. $B$- and $V$-band photometry is relative to the field calibration by \citet{h04} while the $H$-band photometry is relative to 2MASS.}
\label{tab:grb041006_sso}
\end{deluxetable}

\clearpage

\begin{deluxetable}{cccc}
\tablecaption{GRB\,041006 Spectroscopic Lines}
\tablewidth{0pt} \tablehead{ 
\colhead{Observed Wavelength} & \colhead{Equivalent Width} & \colhead{Line ID} & 
\colhead{Rest Wavelenth} \\
\colhead{(\AA)} & \colhead{(\AA)} & \colhead{} & 
\colhead{(\AA)} \\}

\startdata 
4798.16   &      4.4  &   \ion{Mg}{2} & $\lambda2796$ \\
4810.35   &      3.9  &   \ion{Mg}{2} & $\lambda2803$ \\ 
6396.31   &      1.7  &   \ion{O}{2} & $\lambda3727$ \\
6750.32   &      1.4  &   \ion{Ca}{2} & $\lambda3935$ \\
6810.06   &      1.1  &   \ion{Ca}{2} & $\lambda3970$ \\
\enddata 
\label{tab:grb041006_spec}
\end{deluxetable}

\clearpage

\begin{deluxetable}{llrlrl}
\tablecaption{Peak Magnitudes and $^{56}$Ni
  Masses for GRB- and XRF-associated SNe} 
\tablewidth{0pt} \tablehead{ 
\colhead{GRB Name} & \colhead{Redshift} & \colhead{$M_{V,\rm peak}$\tablenotemark{a}} & 
\colhead{Ref.} & \colhead{$^{56}$Ni Mass} & \colhead{Ref.} \\ 
\colhead{} & \colhead{($z$)} & \colhead{(mag)} & \colhead{} &  
\colhead{($M_{\odot}$)} & \colhead{}\\}

\startdata
GRB\,970228 & 0.695 & $-18.16^{+0.54}_{-0.52}$ & \citet{zkh04} & \nodata & \nodata \\
GRB\,980425 & 0.0085 & $-19.13\pm 0.05$ & \citet{gvv+98} & $0.6\pm 0.1$ & \citet{imn+98}, \\
 & & & & & Woosley {\it et al.} (1999) \\
GRB\,980703 & 0.966 & $-19.68^{+1.44}_{-0.60}$ & \citet{zkh04} & \nodata & \nodata \\
GRB\,990712 & 0.434 & $-18.02^{+0.23}_{-0.19}$ & \citet{zkh04} & \nodata & \nodata \\
GRB\,991208 & 0.706 & $-18.86^{+0.72}_{-0.29}$ & \citet{zkh04} & \nodata & \nodata \\
GRB\,000911 & 1.058 & $-18.26^{+0.12}_{-0.14}$ & \citet{zkh04} & \nodata & \nodata \\
GRB\,010921 & 0.45 & $<-17.80$ & \citet{pks+03} & \nodata & \nodata \\
GRB\,011121 & 0.36 & $-18.83^{+0.07}_{-0.10}$ & \citet{zkh04} & \nodata & \nodata \\
GRB\,020405 & 0.698 & $-18.63^{+0.19}_{-0.14}$ & \citet{zkh04} & \nodata & \nodata \\
XRF\,020903 & 0.251 & $-18.53\pm 0.5$ & \citet{skf+05} & \nodata & \nodata \\
GRB\,021211 & 1.006 & $-18.42^{+1.15}_{-0.55}$ & \citet{zkh04} & \nodata & \nodata \\
GRB\,030329 & 0.169 & $-18.83\pm 0.30$ & \citet{dtm+05,log+04} & $0.40^{+0.15}_{-0.10}$ & \citet{dtm+05} \\
GRB\,031203 & 0.1055 & $-19.63\pm 0.15$ & \citet{mtc+04} & \nodata & \nodata \\
XRF\,040701 & 0.2146 & $<-15.95$ & \citet{skf+05} & \nodata & \nodata \\
GRB\,040924 & 0.859 & $-17.63\pm 0.10$ & this paper & \nodata & \nodata \\
GRB\,041006 & 0.716 & $-18.83\pm 0.05$ & this paper & \nodata & \nodata \\

\enddata \tablenotetext{a}{Assuming a SN\,1998bw-like light-curve with no additional stretching.}
\label{tab:grb_MV_Ni}
\end{deluxetable}

\clearpage

\begin{deluxetable}{lrlrl}
\tablecaption{Peak Magnitudes and $^{56}$Ni
  Masses for Nearby SNe Ibc} 
\tablewidth{0pt} \tablehead{ 
\colhead{SN Name} & \colhead{$M_{V,\rm peak}$} & 
\colhead{Ref.} & \colhead{$^{56}$Ni Mass} & \colhead{Ref.} \\ 
\colhead{} & \colhead{(mag)} & \colhead{} &  
\colhead{($M_{\odot}$)} & \colhead{}\\}

\startdata
SN\,1983N & $-18.89\pm 0.57$\tablenotemark{a} & \citet{cwb+96} & 0.25 & \citet{gcd+86} \\
SN\,1983V & $-18.61\pm 0.41$ & \citet{cwp+97} & \nodata & \nodata \\
SN\,1984L & $-18.50\pm 0.10$ & \citet{t87} & $0.20\pm 0.05$ & \citet{byb93} \\
SN\,1987M & $-19.40\pm 0.5$ & \citet{fps90} & $0.26$ & \citet{sfn+93} \\
SN\,1990B & $-17.91\pm 0.3$ & \citet{csp+01} & \nodata & \nodata \\
SN\,1991D & $-19.6\pm 0.6$ & \citet{bbt+02} & 0.7 & \citet{bbt+02} \\
SN\,1992ar & $-19.7\pm 0.5$ & \citet{cps+00} & $0.5^{+0.25}_{-0.18}$ & \citet{cps+00} \\
SN\,1994I & $-17.69\pm 0.58$\tablenotemark{b} & \citet{rvh+96} & $0.07\pm 0.035$ & \citet{inh+94} \\
SN\,1997ef & $-17.1\pm 0.2$ & \citet{inn+00} & $0.15\pm 0.03$ & \citet{inn+00} \\
SN\,1999as & $-21.4\pm 0.2$ & \citet{hbn+01} & $4.0\pm 0.5$\tablenotemark{\dagger} & \citet{hbn+01} \\
SN\,1999ex & $-17.86\pm 0.26$ & \citet{shs+02} & 0.16 & \citet{hsp+03} \\
SN\,2002ap & $-17.4\pm 0.4$\tablenotemark{c} & \citet{fps+03a} & $0.07\pm 0.02$ & \citet{mdm+02} \\
SN\,2003L & $-18.18\pm 0.2$ & Soderberg {\it et al.}, (in prep) & \nodata & \nodata \\
\enddata 
\tablenotetext{a}{Absolute magnitude brightened by 1.22 mag to be in agreement with the distance assumed by \citet{gcd+86}.}
\tablenotetext{b}{Absolute magnitude dimmed by 0.4 mag to be in agreement with the distance assumed by \citet{inh+94}.}
\tablenotetext{c}{Absolute magnitude brightened by 0.2 mag to be in agreement with the distance assumed by \citet{mdm+02}.}
\tablenotetext{\dagger}{Given the similarity of the SN\,1999as light-curve to Type IIn events, it has been suggested that circumstellar interaction contributes to the bright peak luminosity observed for this SN. While there is still no direct indication of such interaction from optical spectra (no narrow emission lines), it is estimated that the explosion must have produced at least $\sim 1~\rm M_{\odot}$ even after correcting for any CSM interaction (Daniel Kasen, PhD thesis).}
\label{tab:sn_MV_Ni}
\end{deluxetable}

\clearpage

\clearpage

\clearpage

\clearpage

\clearpage

\clearpage

\end{document}